\documentclass[11pt,twoside]{article}
\usepackage{asp2010}

\resetcounters

\begin{document}

\title{The shape of dark matter halo in the PRG NGC 4262}
\author{Sergei Khoperskov,$^{1,2}$ Alexei Moiseev,$^{3,2}$ Alexander Khoperskov,$^4$ and Anna Saburova$^2$
\affil{$^1$Institute of Astronomy, Russian Academy of Sciences, Pyatnitskaya st., 48, 119017 Moscow, Russia}
\affil{$^2$Sternberg Astronomical Institute, Moscow MV Lomonosov State University, Universitetskij pr., 13, 119992 Moscow, Russia}
\affil{$^3$Special Astrophysical Observatory, Russian Academy of Sciences, Nizhnii Arkhyz, Karachai-Cherkessian Republic, 357147, Russia}
\affil{$^4$Volgograd State University, Universitetsky pr., 100, 400062 Volgograd, Russia}
}

\begin{abstract}
With the aim to determine the spatial distribution of dark matter, we investigate the polar ring galaxy NGC~4262. We used the stellar kinematics data for the central galaxy obtained from optical spectroscopy together with information about the kinematics of the neutral hydrogen for polar component. It was shown that NGC~4262 is the classic polar ring galaxy case with the relative angle of $88^{\circ}$ between components. From simulations of the central galaxy and ring kinematics we found that the shape of the dark matter distribution varies strongly with the radius. Namely, the dark matter halo is flattened towards the galactic disk plane $c/a=0.4$, however it is prolate to the orthogonal (polar) plane far beyond the central galaxy $c/a=1.7$. Also, the simulations of the ring evolution let us to confirm the stability of the ring and the formation of quasi-spiral structures within it.
\end{abstract}

\section{Introduction}
It was believed that galactic polar structures are the result of close interaction of galaxies, accretion of companion's matter or even cold gaseous filaments. In any case, polar ring galaxies~(PRGs)  appear to be a  good probe to study the dark and baryonic matter gravitational potential around them. The general idea is that polar component should  be stabilized by the dark matter (DM). Rotating in two perpendicular planes gives the opportunity to measure the spatial distribution of the gravitating matter in 3D due to  the decomposition of rotation. Measurements of the kinematics and amount of the baryonic matter  allowed us to constrain the mass and shape of  the dark matter  halo. Theoretical predictions pointed out that halo should be flattened towards the polar plane~\citep{1986MNRAS.219..657S,1982ApJ...263L..51S}. Numerous investigations of the individual  PRGs agree with that \citep[e.g.][]{1993A&A...267...21A,1996A&A...305..763C,2003ApJ...585..730I}. However, the amount of real objects with the well-measured halo axis ratio is still limited. Based on new observations we decided to enlarge the sample of the known PRGs with measured shape parameters of the DM. In this paper we present   new results, which are obtained for NGC~4262.

\subsection{Observations}
\begin{figure*}
\centerline{
\includegraphics[width=0.5\textwidth]{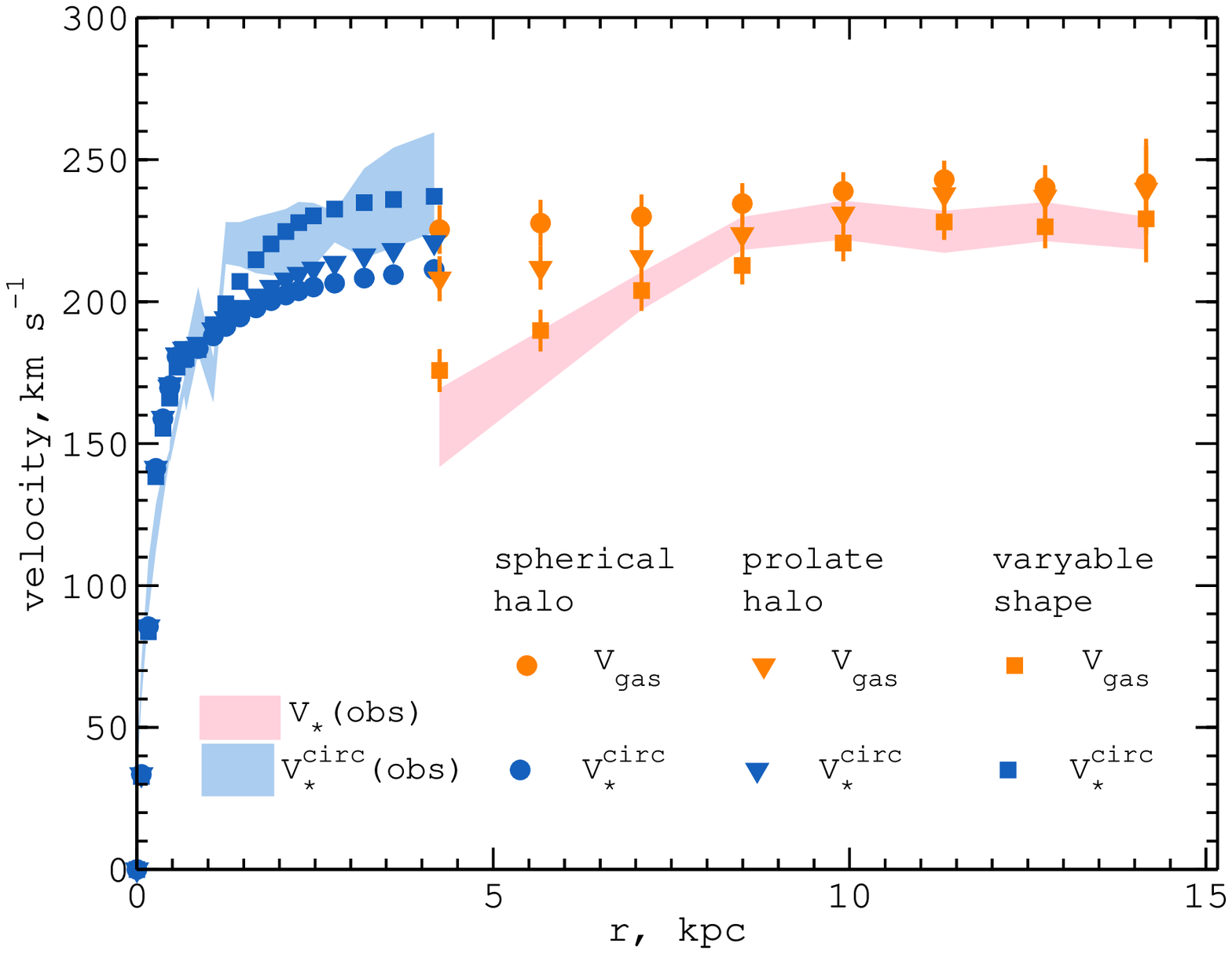}
\includegraphics[width=0.5\textwidth]{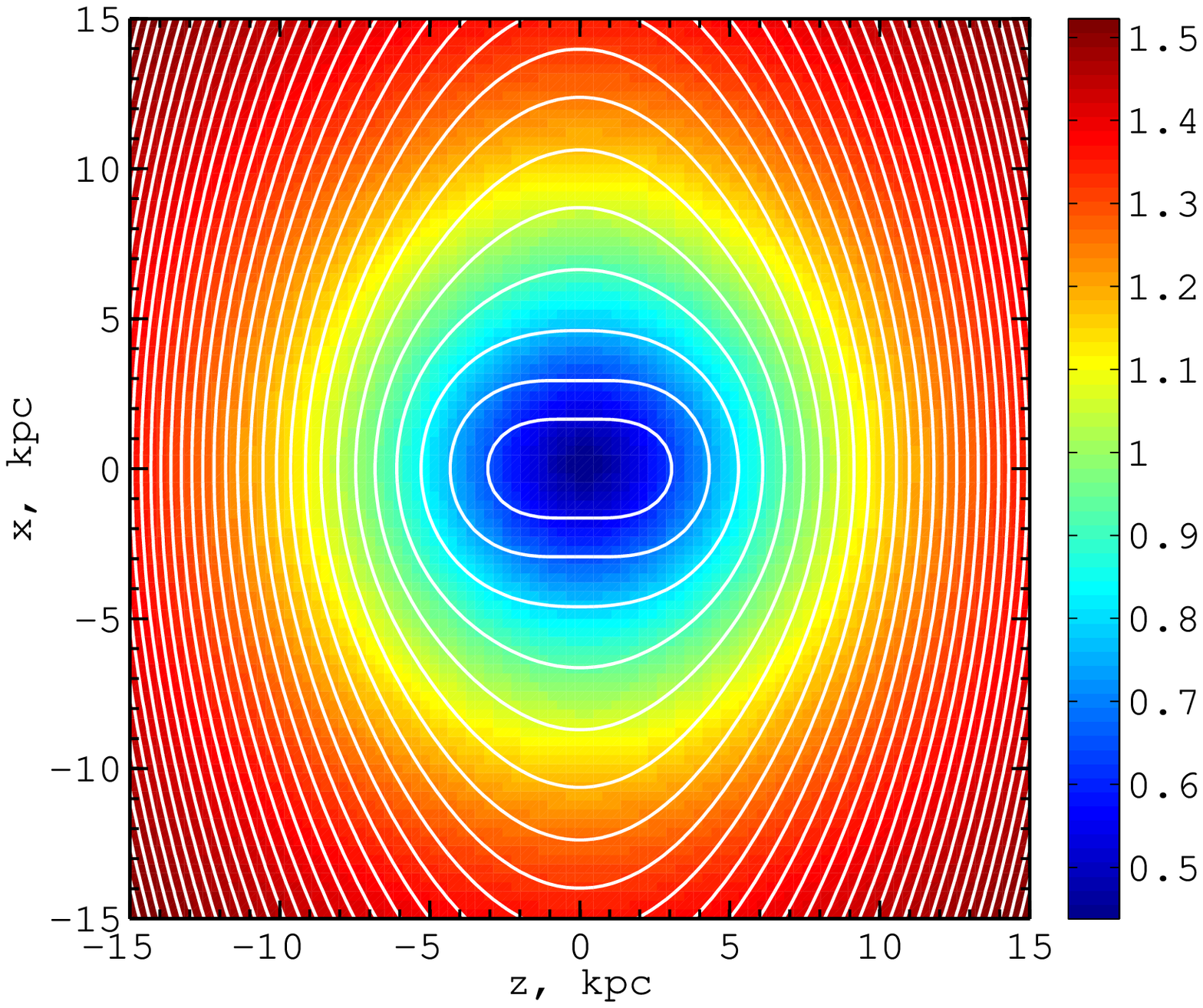}
}
 \caption{Left plot: observed kinematics of NGC~4262 (filled) and  best-fit models for the spherical halo (circle), prolate halo (triangle) and variable shape halo (square). Right plot: contours of the halo potential for the best-fit model. Color represents the value of the halo axis ratio.}
\label{fig_1}
\end{figure*}

The central object is the early-type barred galaxy~\citep{SAURONIII}. The external wide HI ring beyond the stellar disk of the galaxy was discovered by~\citet{1985A&A...144..202K}. \citet{Bettoni2010} showed that a small contribution of stellar population observed in UV bands is presented in the ring, which  is significantly inclined or even polar to the central disk \citep[see also the comments in][]{2011Ap&SS.335..231B}. The galaxy seems to be an ideal object for our simulations because the polar ring mostly consists  of the gaseous component with well-determined surface mass density distribution and kinematics. We used new 21-cm WSRT observations presented in  \citet{2010MNRAS.409..500O} to  calculate the gas density distribution and rotation curve of the ring as well. Elliptic beam was taken into account for the rotation curve reconstruction (see figure~\ref{fig_1}, left) with the help of the TiRiFiC software~\citep{TiRiFiC2007}. Rotation curve and velocity dispersion of the CG were derived from the long-slit observations with the multi-mode focal reducer SCORPIO-2~\citep{2011BaltA..20..363A} at the SAO RAS 6-m BTA telescope.  Analysis of kinematics and morphology properties of CG and PR gives us the mutual angle between the components $\delta = 50 \pm 6\deg$ or $\delta = 88 \pm 6\deg$. Thus, the second value corresponds to the classic polar ring galaxy case.

%%%%%%%%%%%%%%%%%%%%%%%%%%%%%%%%%%%%%%%%%%%%%%%%%%%%%%%%%%%%%%%%%%%%%%%
%%%%%%%%%%%%%%%%%%%%%%%%%%%%%%%%%%%%%%%%%%%%%%%%%%%%%%%%%%%%%%%%%%%%%%%

\subsection{Halo model}
Our  kinematical  models of both the CG and PR are based on the assumption of the stability of the ring. We simulated the line-of sight velocity field of the PR and circular velocity of the CG simultaneously. The circular velocity for CG (figure~\ref{fig_1}, left) was calculated  from the asymmetric drift equation using the information about stellar velocity dispersion distribution along a radius.  The isothermal halo density profile was used in the simulations. Three general types of the DM halo shape were tested: spherical halo, oblate halo and variable shape halo. Using $\chi^2$ minimization we showed that  the  models with spherical and prolate halo do not reproduce well the observed kinematics of NGC~4262~(see figure~\ref{fig_1}).  In these models the simulated circular velocity of the CG is systematically lower than observed. Vice versa gaseous rotation  amplitude in the polar plane is a bit higher than observed, especially in the inner  region of the ring. Thus, to achieve the better agreement with observations we investigate the model with variable shape of DM potential.
For the gravitational potential the following axis ratio model was accepted~\citep{2007MNRAS.377...50H}:
\begin{equation}
\displaystyle c/a = \exp{ \left[ \alpha \tanh{\left( \gamma \log{\left(r/r_{\alpha}\right)}\right)}\right]}\,,
\end{equation}\label{eq::c_to_a_by_Hayashi}
where parameters $\alpha$, $\gamma$ and $r_{\alpha}$ determine the character of the halo shape.
The best fit model implies the strong transformation of the halo axis ratio with radius. Better agreement between simulated and observed kinematics was found for  $\alpha = 0.9 \pm 0.3$, $\gamma = 0.68 \pm 0.09$ and $r_{\alpha} = 1.33 \pm 0.2$.  The result is clearly seen in the left panel of figure~\ref{fig_1}. The central part of the halo (within $r < 5$~kpc) is flattened towards the CG plane: $c/a = 0.4$. At $r \approx 5$~kpc it is round, and far beyond CG the halo is flattened towards the polar plane $c/a = 1.54$ at $r=15$~kpc. In figure~\ref{fig_1} (right)  we show  both the contours of the potential and $c/a$ colormap. Smaller central value of the $c/a$ gives  low halo spatial scale leading to the faster increasing of the circular velocity of CG. At the same time high effective halo scale in the polar plane gives the slow growth of the rotation velocity for the polar component.

Despite the variable shape of DM around galaxies is expected from the different numerical simulations \citep[e.g.][]{2012MNRAS.425.1967S}, there is not  much evidence of such halo type in galaxies. The first evidences of the variable halo shape were obtained for the Milky Way  from the analysis of the stellar streams~\citep{2013ApJ...773L...4V}. Flaring of the gas layer at the outskirts of the Galaxy also agrees with the non-constant halo axis ratio~\citep{2011ApJ...732L...8B}. The question is whether the halo consists only of the collisionless matter or  it includes also the dark baryonic gas. Unfortunately, within the current work there is no opportunity to separate  dissipative and baryonic fraction of the NGC~4262 halo.

%%%%%%%%%%%%%%%%%%%%%%%%%%%%%%%%%%%%%%%%%%%%%%%%%%%%%%%%%%%%%%%%%%%%%%%
\begin{figure*}
\includegraphics[width= \textwidth]{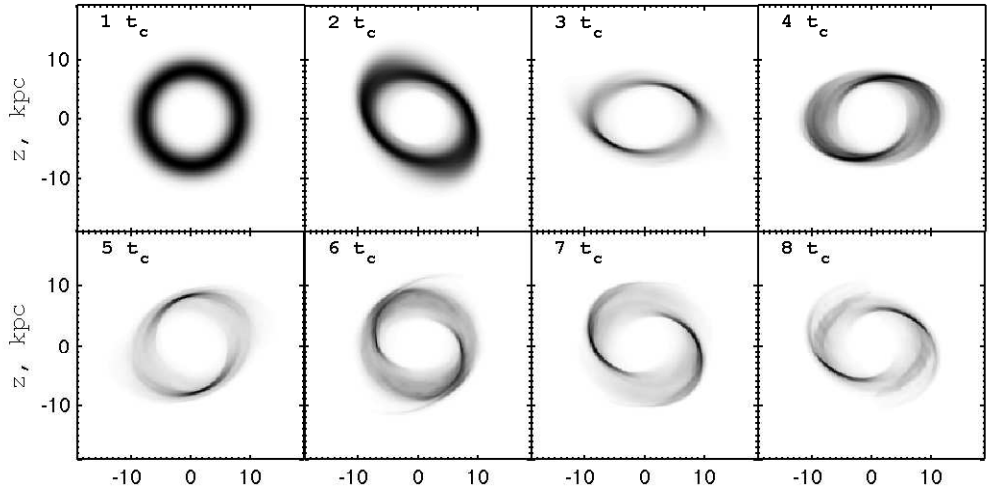}
 \caption{Time dependent evolution of the column density of the gaseous polar component.}
\label{fig_2}
\end{figure*}
%%%%%%%%%%%%%%%%%%%%%%%%%%%%%%%%%%%%%%%%%%%%%%%%%%%%%%%%%%%%%%%%%%%%%%%

\subsection{Dynamical simulations}
Dynamical simulations are the important tool to test the stability of the polar component within the complicated gravitational potential obtained above. Our model is based on the hydrodynamical simulations of the evolution of the gaseous ring which is submerged into the massive dark matter halo. Simulations were performed using grid-based TVD MUSCL scheme. The initial conditions  correspond to the best-fit model of the variable DM shape from previous paragraph.  We neglected the contribution of the ring component to the total gravitational potential, because its total stellar and gaseous masses were low.
Figure~\ref{fig_2} shows the time dependent evolution of the column density of the gas in the polar plane. The polar ring is stable during numerous periods of rotation. However, the non-axisymmetric perturbation from the halo generates the spiral structure in the gaseous polar ring. The influence of the DM shape is larger than the perturbation from the disk at the distances of $5-15$~kpc from the center. Global spiral pattern could lead to the formation of the dense regions where the star formation processes should be driven.
These morphological features are the similar to the arc-like (or spiral fragment) structure with the ongoing star formation in the nothern-west side of the polar ring of NGC~4262 detected on the \textit{GALEX} UV images \citep{Bettoni2010}.

\acknowledgements This work was supported by the RFBR grants 12-02-00685, 12-02-31452, 13-02-00416 and the ``Active Processes in Galactic and Extragalactic Objects'' basic research program of the Department Physical Sciences of the RAS OFN-17. S.K., A.M. and A.S. are also grateful for the financial support of the `Dynasty' Foundation. The observations at the 6-m telescope were carried out with the financial support of the Ministry of Education and Science of Russian Federation (contracts no. 16.518.11.7073 and 14.518.11.7070).

%\bibliography{editor}
\bibliography{aspauthor}

\end{document}